\documentclass[aps,prb,twocolumn]{revtex4-2}
\usepackage{color}
\usepackage{graphicx}
\usepackage{cancel}
\usepackage{amsfonts}
\usepackage{amssymb}
\usepackage{amsthm}
\usepackage{ulem}
\usepackage{url}
\usepackage{hyperref}
\usepackage{float}
\usepackage{bm}        
\usepackage{enumerate}
\usepackage{mathtools}

\usepackage[dvipsnames]{xcolor}
\hypersetup{
	colorlinks,
	linkcolor={green!50!black},
	urlcolor={blue!80!black},
	citecolor   = {red!50!black} 
}

\usepackage{amsmath}
\usepackage{braket}
\usepackage{color}
\usepackage{bm}
\usepackage{bbold}
\usepackage{graphicx}
\usepackage{mathtools}
\usepackage{bigints}
\usepackage{float}

\usepackage{soul}

\def\bv{{\bf B}}
\def\ev{{\bf E}}

\def\n{{\bf n}}

\begin{document}

\title{Skyrmion-driven topological spin and charge Hall effects in diffusive antiferromagnetic thin films}

\author{Amir N. Zarezad}
\author{Józef Barnaś}
\author{Anna Dyrda\l}
\email[e-mail:]{anna.dyrdal@amu.edu.pl}
\affiliation{Department of Mesoscopic Physics, ISQI, Faculty of Physics,
Adam Mickiewicz University in Poznań,\\ ul. Uniwersytetu Poznanskiego 2, 61-614 Pozna\'n, Poland}

\author{Alireza Qaiumzadeh}
\affiliation{Center for Quantum Spintronics, Department of Physics, Norwegian University of Science and Technology, NO-7491 Trondheim, Norway}

\begin{abstract}
We investigate topological Hall effects in a metallic antiferromagnetic (AFM) thin film and/or at the interface of an AFM insulator--normal metal bilayer with a single skyrmion in the diffusive regime. 
To determine the spin- and charge Hall currents, we employed a Boltzmann kinetic equation with both spin-dependent and spin-flip scatterings. The interaction between conduction electrons and static skyrmions is included in the Boltzmann equation {\it via} the corresponding emergent magnetic field arising from the skyrmion texture. We compute intrinsic and extrinsic contributions to the topological spin Hall effect and spin accumulation, induced by an AFM skyrmion. We show that although the spin Hall current vanishes rapidly outside the  skyrmion, the spin accumulation can be finite  at the edges far from the skyrmion, provided the spin diffusion length is longer than the skyrmion radius. 
In addition, We show that in the presence of a spin-dependent relaxation time, the topological charge Hall effect is finite and we determine the corresponding Hall voltage. 
Our results may help to explore antiferromagnetic skyrmions by electrical means in real materials. 
\end{abstract}

\date{\today}

\maketitle

Interplay between charge currents and magnetic textures is important from both fundamental and applied points of view ~\cite{PhysRevLett.78.3773, PhysRevLett.79.5110, PhysRevB.59.138, PhysRevB.60.3406, PhysRevB.63.224412, 10.1063/1.1578165, Yamanouchi2004, PhysRevLett.92.086601, TATARA2008213, Swaving_2012, PhysRevLett.110.127208, PhysRevB.93.180408, PhysRevB.96.165303, PhysRevB.102.184413}.
Interactions between current and ferromagnetic skyrmions may lead to emerging phenomena, including topological Hall and skyrmion Hall effects ~\cite{Liang2015, WANG2022100971}.  
The spin-polarized charge current flowing in a ferromagnetic (FM) layer containing a topological skyrmion drags it along the electric field and simultaneously deflects it toward one of the film edges, that is, along the direction perpendicular to the electric field under the influence of gyrotropic forces. The latter phenomenon
is called the skyrmion Hall effect~\cite{nphys2231, nnano.2013.243, ncomms2442, Litzius2017, nphys3883, PhysRevB.95.094401}. This phenomenon was studied theoretically and observed experimentally; see, e.g. Refs.~\cite{Litzius2017, nphys3883, PhysRevLett.100.127204, PhysRevB.78.134412, PhysRevLett.107.136804, Hans2012, Tomasello2014} for an overview.  The motion of rigid topological skyrmions in both longitudinal and transverse directions in the present of charge currents in FM metals are conveniently described by Thiele's equation~\cite{ncomms2442, PhysRevLett.30.230, Martinez_2016}. 
In turn, FM skyrmions affect the flow of spin-polarized charge currents in FM metal by deflecting their trajectories in the direction perpendicular to the external electric field~\cite{PhysRevLett.83.3737, PhysRevLett.93.096806, PhysRevLett.102.186602, PhysRevLett.102.186601, PhysRevB.91.245115, PhysRevLett.117.027202, PhysRevB.95.064426,PhysRevB.97.134401}. This is a phenomenon similar to the anomalous Hall effect in FM metals with uniform magnetization direction \cite{ PhysRev.95.1154, RevModPhys.82.1539, PhysRevLett.112.017205}. The origin of this skyrmion-induced topological Hall effect is the emergence of a real-space Berry curvature, induced by skyrmion textures ~\cite{ PhysRevLett.93.096806, PhysRevLett.98.246601}, while in the anomalous Hall effect, the Berry curvature in the momentum space arises from the spin-orbit couplings ~\cite{ PhysRevB.64.104411, PhysRevB.64.014416, PhysRevB.68.045327}. The FM skyrmion Hall effect and the topological Hall effect of spin-polarized currents in FM metals are reciprocal phenomena with a similar origin.

On the other hand, AFM skyrmions, unlike their FM counterparts, do not show any skyrmion Hall effect \cite{APL1.4967006, PhysRevLett.116.147203, Zhang2016_1, Zhang2016, PhysRevB.99.054423, TRETIAKOV2021333, GOBEL20211, Amin2023}. This deflection-free motion of AFM skyrmions is one of the advantages of AFM systems versus their FM counterparts in practical spintronic devices.  
However, topological skyrmion textures may still create a real-space Berry curvature \cite{PhysRevB.86.245118} and hence some type of topological Hall effects are expected to be present in AFM systems in the presence of skyrmions.
Charge currents in compensated AFM metals are not spin-polarized, and thus the two equally populated opposite spins can be deflected in opposite directions in the presence of the AFM skyrmion-induced Berry curvature.
Therefore, we can expect the topological \emph{spin} Hall effect instead of the topological charge Hall effect.
Recently, this effect has been investigated theoretically in a few publications~\cite{RRL1700007, PhysRevB.96.060406, PhysRevLett.121.097204, nakazawa2023topological}, but has not been measured experimentally yet. The absence of a topological charge Hall effect was recently confirmed in a chiral AFM system \cite{Qiang_2022}.

In Ref. \cite{RRL1700007}, the authors employed an SU(2) semiclassical framework in combination with the ab-initio description of the electronic structure and demonstrated the emergence of a sizeable transverse spin current and a vanishing transverse charge current in 
a synthetic AFM skyrmion lattice. 
In Ref. \cite{PhysRevB.96.060406}, the authors showed a finite topological spin Hall effect and the absence of topological charge Hall effect in a compensated AFM skyrmion crystal on a honeycomb lattice.
Akosa {\it et al}~\cite{PhysRevLett.121.097204} computed the topological spin and charge Hall effects in a finite size AFM square lattice using a tight-binding model in terms of Landauer–B\"{u}ttiker formalism, implemented in the Kwant code, in the presence of electrostatic impurities. They found a zero topological charge Hall effect and a nonzero topological spin Hall effect in such systems. Finally, in Ref. \cite{nakazawa2023topological}, using Landauer–B\"{u}ttiker formula, the authors showed that a vector chirality, formed by the AFM N{\'e}el vector, gives rise to a finite topological spin Hall effect in bulk system of AFM half-skyrmions or merons but not in skyrmions.

In the present article, we investigate the topological charge and spin Hall effects in a diffusive AFM layer with a skyrmion texture, by means of the quasiclassical Boltzmann kinetic equation. 
We consider both spin-dependent and spin-flip scatterings in our formalism, and we find analytical expressions for transverse spin and charge currents as well as spin accumulation in the presence of an AFM skyrmion. 
We show that spin-dependent scatterings generate a nonzero topological charge Hall effect. In addition, we compute both intrinsic (disorder independent, which is arising from the real-space Berry curvature) and extrinsic (disorder dependent, which is arising from the spin-flip scatterings) contributions to the topological spin Hall effect and spin accumulation.

It is worth mentioning that detecting AFM topological textures, such as skyrmions and merons, is a challenging experimental task. Only recently has some evidence of AFM skyrmions and merons been reported in bulk AFM systems and synthetic AFM multilayers~\cite{Gao,Jani,Legrand}.
Our findings on charge and spin Hall effects in AFM skyrmions can be used to explore these topological textures in experiments.

\section{Model}
We consider a two-sublattice compensated AFM square lattice that hosts a single static skyrmion. The system consists of an electronic subsystem and a AFM spin subsystem that interact to each other via an sd exchange interaction. The effective Hamiltonian of the system can be written in the following form~\cite{PhysRevLett.121.097204,   PhysRevB.95.134424, PhysRevLett.120.197202, PhysRevResearch.5.L022065}:  
\begin{equation}
\hat{H} = -t \gamma_{\mathbf{k}}\big(\hat{\tau}_x\otimes\hat{s}_{0}\big)-J \big(\hat{\tau}_z\otimes\bf {n}\big)\cdot\hat{\bf{s}}.
\label{Hamiltonian}
\end{equation}
Here, ${t>0}$ is the hopping parameter, ${\gamma_{\mathbf{k}}=z^{-1}\sum_{i=1}^z \exp{(-i \bm{k} \cdot \bm{\delta}_i)}}$ is the lattice structure factor with $\bm{\delta}_i$ denoting the nearest-neighbor vectors and $z$ being the coordination number. For a square lattice ($z$=4), we have $\gamma_{\mathbf{k}}=\big(\cos (k_xa)+\cos (k_ya)\big)/2$, where $a$ is the lattice constant. Furthermore, $J$ is the sd exchange interaction that parameterizes the coupling of mobile electrons and localized magnetic moments, $\hat{\bm{\tau}}$ and $\hat{\bm{s}}$ are the vectors of Pauli matrices representing
the AFM sublattice and spin degrees of freedom, respectively, $\hat{s}_0$ is the identity matrix in the spin subspace, and
$\mathbf{n}$ is the unit vector of the staggered  N{\'e}el order parameter field.
In general, the unit vector of the staggered order parameter field can be decomposed into two terms, $\mathbf{n}=\mathbf{n}_0+\mathbf{n}_r$, where $\mathbf{n}_0$ represents the background homogeneous field while the second term $\mathbf{n}_r$ represents the noncollinear magnetic texture~\cite{PhysRevB.98.195439}.

The corresponding eigenvalues and eigenstates of the AFM Hamiltonian in the absence of any texture, $\bf n_r=0$, are given by ~\cite{PhysRevLett.121.097204,PhysRevB.95.134424};
\begin{equation}
\varepsilon_{\eta}(\mathbf{k})=\eta\sqrt{t^2\gamma_{\mathbf{k}}^2+J^2},
\label{eigenvlue}
\end{equation}
 \begin{equation}
\ket{\Psi_{\eta}^{s}}=\dfrac{1}{\sqrt{2}}\Big(\sqrt{1+s\eta \mathrm{P}_{\mathbf{k}}}\ket{A}+\eta
\sqrt{1-s\eta \mathrm{P}_{\mathbf{k}}}\ket{B}\Big)\otimes\ket{\bf \sigma},
\label{eigenfunction}
\end{equation}
where, $\eta=+1$ ($-1$) corresponds to the conduction (valence) band, and $s=+1$ ($-1$) corresponds to the spin up
(down) state, 
$\ket{A (B)}$ refers to the AFM sublattice A (B) projection, $\ket{\sigma}=\ket{\uparrow (\downarrow)}$ denotes up (down) spin projection, and finally we defined  $\mathrm{P}_{\mathbf{k}}=J/\sqrt{t^2\gamma_\mathbf{k}^2+J^2}$. 

We consider a compensated AFM system that preserved combined time and inversion symmetry ($PT$ symmetry) and thus the electronic band dispersion, Eq. (\ref{eigenvlue}), is spin-degenerate. 
To find analytical results, we assume the Fermi level near the maximum of the conduction (minimum of the valence) band, and consider the electronic dispersion around the $\Gamma$ point. Accordingly,  the structure factor is approximated as $\gamma_{\mathbf{k}} \approx (1-a^2k^2/4)$, while the energy dispersion around the $\Gamma$ point becomes $\varepsilon_\eta \approx \eta \big(\sqrt{J^2+t^2}-\hbar^2 k^2/(2m_{\rm eff}) \big)$. The first term here is the energy at the $\Gamma$ point and the second term is the effective kinetic energy of electrons with an effective mass $m_{\rm eff}=2\hbar^2\sqrt {J^2+t^2} /a^2t^2$.

The profile of an AFM skyrmion  can be modelled in the spherical coordinates as,
\begin{equation}
\label{eq:n}
\mathbf{n}_r = \big(\cos{\Phi} \sin{\Theta}, \sin{\Phi} \sin{\Theta}, \cos{\Theta} \big),
\end{equation}
where the polar $\Theta(r)$ and azimuthal $\Phi(r)$ angles are defined as~\cite{PhysRevB.95.064426, PhysRevLett.121.097204} 
\begin{subequations}\label{profile}
\begin{align}
\Theta  &= 2\pi - 4\, \arctan \Bigl(\exp(4r/r_{\rm sk})\Bigr), \\
\Phi &= q \rm{Arg}(x + i y) + c \frac{\pi}{2}.
\end{align}
\end{subequations}
Here, $r=\sqrt{x^2+y^2}$ is the distance from the skyrmion center, $r_{\rm sk}$ denotes the radius of the skyrmion core,  
 $p = \pm 1$ stands for the skyrmion polarity, $q = \pm 1$ denotes skyrmion vorticity, and $c = \pm 1$ defines  the chirality of the skyrmion. 

For a spin texture slowly varying in space, the exchange interaction term -- the second term in the Hamiltonian (\ref{Hamiltonian}) -- can be diagonalized by a unitary gauge transformation. 
The result is a uniform spin background in the presence of an emerging SU(2) gauge field that interacts with itinerant electrons~\cite{TATARA2008213, PhysRevB.86.245118,PhysRevLett.98.246601, Volovik1987, PhysRevB.77.134407, PhysRevLett.102.086601}. The emergent magnetic field  induced by an AFM texture depends on the spin, band, and sublattice indices \cite{PhysRevLett.121.097204, akosa1},
\begin{equation}
B^{\alpha,s}_{\mathrm{em},\eta}(\bm{r})=-s\big(1+s\eta\alpha \mathrm{P}_{\mathbf{k}}\big)\dfrac{\hbar}{2e}N_{x,y}(\bm{r})\hat{z},\\[8pt]
\label{eq:emergent_field}
\end{equation}
where $\alpha=+(-)$ refers to the $A (B)$ sublattice, $\hbar=h/(2\pi)$ is the reduced Planck constant, $e$ is the electron charge, and $N_{x,y}(\bm{r})=\n\cdot(\partial_x \n\times\partial_y \n)$ is the topological charge density. 
Note, the above description is based on the continuous model of skyrmion texture. Such a description relies on the assumption  that variation of the  Neel vector on the atomic distance is small, which is generally fulfilled as the skyrmion size is much longer than the interatomic distance. An alternative description of skyrmions is based on atomistic simulation approach. However, we employ the commonly used continuous model as it allows to achieve some analytical formulas.   

Without loss of generality, we assume the Fermi level is located in the conduction band and in the rest of this article, we set $\eta=+1$ and drop this index.
We consider a narrow AFM stripe of width $2w \geq 2r_{\rm sk}$ and length $2L$, that includes a single skyrmion at its center.

\section{Boltzmann kinetic equation}
To describe spin and charge transports in the diffusive regime in the presence of emerging magnetic field induced by a static AFM skyrmion, we employ a semiclassical transport theory based on the Boltzmann kinetic equation~\cite{Ashcroft, PhysRevB.67.052407, PhysRevB.82.184423}, 
\begin{eqnarray}
\label{eq:BEq}
        \mathbf{v_k} \cdot \frac{\partial f_{s}}{\partial \mathbf{r}} - \frac{e}{\hbar} \Bigl(\mathbf{E} + \mathbf{v_k} \times \mathbf{B}_{\mathrm{em}}^{s} \Bigr) \cdot \frac{\partial f_{s}}{\partial \mathbf{k}} =  
- \frac{f_{s} - \langle f_{s}\rangle}{\tau_{s}} \nonumber \\- \frac{\langle f_{s}\rangle - \langle f_{-s}\rangle}{\tau_{\rm sf}},
\end{eqnarray}
where $f_{s} = f_{s}(\mathbf{r}, \mathbf{k})$ is the nonequilibrium distribution function for electrons with spin $s = +/- $ (or equivalently $s=\uparrow/\downarrow$), $\bf v_k$ is the electron velocity, $\mathbf{E}=E_x \hat{x}$ is the external electric field applied along the stripe, $\langle f_{s} \rangle = \int d^{2}\Omega_{\mathbf{k}} f_{s} / \int d^{2}\Omega_{\mathbf{k}}$ is the angular average over the momentum space, and $\Omega_{\mathbf{k}}$ is the solid angle in the momentum space. The first term on the right hand side describes the spin-conserving relaxation processes with $\tau_{s=\uparrow (\downarrow)}$ being the corresponding spin-dependent scattering time. The second term describes spin mixing relaxation processes with $\tau_{\rm sf}$ denoting the corresponding spin-flip relaxation time. 
The total spin-dependent emerging magnetic field is the sum of the two sublattices contributions,
\begin{equation}
\mathbf{B}_{\mathrm{em}}^{s}(\bm{r})=\sum_{\alpha=A, B} \bv^{\alpha,s}_{\mathrm{em}}(\bm{r})=s B_{\rm em}(\bm{r})\hat{z}, \label{effective_field}
\end{equation}
where for a skyrmion profile, defined in Eqs. (\ref{eq:n}) and (\ref{profile}), the emerging magnetic field amplitude reads~\cite{nnano.2013.243}, 
\begin{align}
\dfrac{B_{\rm em}}{B_0}=-\frac{8}{r/r_{sk}}\,  \sin\left(4\arctan (\exp(\frac{4r}{r_{\rm sk}}))\right)\frac{\exp(\frac{4r}{r_{\rm sk}})}{1+\exp(\frac{8r}{r_{\rm sk}})},
\label{eq:emergent_field}
\end{align}
with $B_0=(h /e)(\pi r_{\rm sk}^2)^{-1}$. The emergent magnetic field is normal to the AFM layer and has an opposite sign for up and down itinerant electron spins. Therefore, this magnetic field effectively deflects the electron trajectory of spin-up and spin-down electrons in opposite directions. If there is no asymmetry between the spin up and spin down electron subbands, this leads to a vanishing topological charge Hall effect, while the spin Hall effect is nonzero. However, in the presence of an asymmetry between spin subbands, e.g., due to different relaxation times or breaking the time-reversal symmetry of electronic bands, both charge and spin Hall effects may occur, as we will show later.       

\section{Analytical solutions of Boltzmann equation }
To solve the Boltzmann kinetic equation, Eq. (\ref{eq:BEq}), we decompose the nonequilibrium distribution function into an equilibrium component $f_{s}^{0}$ and small nonequilibrium perturbations \cite{PhysRevB.97.134401, valet, 10.1063/1.356868},
\begin{equation}
\label{eq:fs}
f_{s} = f_{s}^{0} - \frac{\partial f_{s}^{0}}{\partial \varepsilon} \Bigl(- e \mu_{s}(\mathbf{r}) + g_{s}(\mathbf{r}, \mathbf{k}) \Bigr),
\end{equation}
where $-e\mu_{s}(\mathbf{r})$ and $g_{s}(\mathbf{r}, \mathbf{k})$ are the isotropic (zeroth velocity moment) and anisotropic (first velocity moment) parts, respectively. The first part is related to spin accumulation, while the second part represents a shift of the electron sphere in momentum space. 
Both $\mu_{s}(\mathbf{r})$ and $g_{s}(\mathbf{r}, \mathbf{k})$ should be determined from the Boltzmann equation. Inserting Eq. (\ref{eq:fs}) into Eg. (\ref{eq:BEq}), and separating the odd and even velocity moments of the distribution function, up to linear order in the emerging magnetic field, we find in the zero-temperature  limit \cite{PhysRevB.97.134401, valet, 10.1063/1.356868},
\begin{align}
&g_s({\mathbf{k}},{\bf r}) = -e\tau_s{\bf v}_{\mathbf{k}}\cdot \left({\bf{E}}-\nabla_{\bf r}\mu_{s}({\mathbf{r}})-\frac{e\tau_s}{m_{\rm eff}} {\bf E} \times \mathbf{B}_{\rm em}^{s}(\bm{r})\right), \label{eq:g3}
\\
&\nabla^2\delta\mu(\mathbf{r})-\dfrac{\delta\mu(\mathbf{r})}{\lambda_{\rm sd}^2}=\dfrac{e\tau}{m_{\mathrm{eff}}}(\hat{z}\times\ev)\cdot\nabla\mathbf{\bm{B}}_{\rm em}(\bm{r}).
\label{eq:diffusion_sub_lattice}
\end{align}
Here, $\delta\mu=(\mu_{\uparrow}-\mu_{\downarrow})/2$ is the net spin accumulation,
$\tau=(\tau_{\uparrow}+\tau_{\downarrow})/2$ is the spin-averaged relaxation time, $\lambda_{\rm sd}$ is the spin diffusion length, defined as $\lambda_{\rm sd}^{-2} =(l_{\uparrow}^{-2} +l_{\downarrow}^{-2})/2$, where $l_{s}^2=v_{\rm F}^2\tau_s\tau_{\rm sf}/2$ and $v_{\rm F}$ is the electron Fermi velocity.  
The difference between two spin-dependent scatterings can be quantified by a spin asymmetry relaxation time parameter $p_\tau= (\tau_\uparrow - \tau_\downarrow )/(\tau_\uparrow + \tau_\downarrow)$. We can also define spin asymmetry of the spin-dependent conductivity as $p_\sigma=(\sigma_\uparrow -\sigma_\downarrow )/
(\sigma_\uparrow +\sigma_\downarrow )$. In AFM metals with degenerate spin bands $p_\tau=p_\sigma$ while in spin nondegenerate bands, such as FM metals, these two parameters can be different \cite{PhysRevB.97.134401}.

The spin  current density is determined from the formula 
${\bf j}_{s}={-e}(2\pi)^{-2}\int d^2\mathbf{k} \,f_{s}(\bf r,\mathbf{k}) {\bf {v}}_{\mathbf{k}}$,
which upon using Eq. (\ref{eq:g3}) leads to the following relation,
\begin{equation}
{\bf j}_{s}(\bm{r})=\sigma_{s}\Big(\ev-\nabla_{\bf r}\mu_{s}(\bm{r})-\dfrac{e\tau_{s}}{m_{\rm eff}}\ev\times\mathbf{B}_{\rm em}^{s}(\bm{r})\Big),
\label{eq:g_s}
\end{equation}
where $\sigma_{s}=(e^2/2h)(v_Fk_F\tau_{s})$ is the spin-dependent conductivity.
The total charge and spin current densities are respectively defined as,
\begin{subequations}\label{currents}
  \begin{align}
{\bf j}^{\rm ch}={\bf j}_{\uparrow}+{\bf j}_{\downarrow},\\
{\bf j}^{\rm sp}={\bf j}_{\uparrow}-{\bf j}_{\downarrow}.
\end{align}  
\end{subequations}

We are interested in the average of the spin accumulation along the transport direction, i.e., $x$ direction. Equation (\ref{eq:diffusion_sub_lattice}) can then be rewritten as, 
\begin{equation}
\dfrac{d^2\overline{\delta\mu}(y)}{dy^2}-\dfrac{\overline{\delta\mu}(y)}{\lambda_{\rm sd}^2}=\dfrac{e\tau E_x}{m_{\rm eff}} \dfrac{d\overline{B}_{\rm em}(y)}{dy},
\label{eq:diffusion_sub_lattice1}
\end{equation}
where we defined $\overline{F}(y)=(2L)^{-1}\int_{-L}^{L} dx F({\bm{r}}) $.
The general solution of this differential equation is the sum of homogeneous and particular solutions. 
The homogeneous solution includes two unknown constants that must be determined from the appropriate boundary conditions. 

Similarly, using Eqs. (\ref{eq:g_s}) and (\ref{currents}), we find the following differential equations for the transverse charge and spin current densities:
\begin{subequations}\label{current1}
\begin{align}
& \overline{j}^{\rm ch}_{y}(y)=-\sigma_0\left(\dfrac{d \overline{\mu}(y)}{dy}+p_{\tau}\dfrac{d \overline{\delta\mu}(y)}{dy}\right)+\dfrac{e \tau\sigma_0}{m_{\rm eff}}
p_{\tau}E_x\overline{B}_{\rm em}(y),\label{eq:j_ch}\\[8pt]
& 
\overline{j}^{\rm sp}_{y}(y)=-\sigma_0\left(p_{\tau}\dfrac{d \overline{\mu}(y)}{dy}+\dfrac{d \overline{\delta\mu}(y)}{dy}\right)\nonumber\\
&\hspace{3cm} +\dfrac{e \tau\sigma_0}{2m_{\rm eff}}\big(1+p^2_{\tau})E_x\overline{B}_{\rm em}(y),\label{eq:j_sp}
\end{align}
\end{subequations}
where $\sigma_0=\sigma_{\uparrow}+\sigma_{\downarrow}$ is the total longitudinal conductivity of the AFM metal and $\mu=(\mu_{\uparrow}+\mu_{\downarrow})/2$ is the spin-averaged chemical potential.

To solve the differential equations (\ref{eq:diffusion_sub_lattice1}) and (\ref{current1}), and to find expressions for the spin accumulation, spin Hall current, and charge Hall voltage in the assumed stripe geometry, we need now to use the appropriate boundary conditions. 
Since the AFM stripe has a finite width with an open boundary condition, the transverse component of the charge current density must be zero, $\overline{j}^{\rm ch}_{y}(y)=0$. Consequently, using Eq.~(\ref{eq:j_ch})  one finds the average transverse electric field and Hall voltage,
\begin{subequations}
\begin{align}
 \overline{E}_y(y)&=-\dfrac{d{\overline\mu} (y)}{dy}=p_{\tau}\left(\dfrac{d\overline{\delta\mu}(y)}{dy}-\dfrac{2e\tau}{m_{\rm eff}}E_x \overline{B}_{\rm em}(y)\right), \label{E_Hall}\\
 V_H&=\int_{-w}^{w} \overline{E}_y(y)dy.
\label{eq:spin_motive_field}
\end{align}
\end{subequations}
As it is evident from the above expressions, the Hall voltage is finite only if there is an asymmetry between the relaxation time of up and down spins, i.e., $p_\tau \ne 0$.
Indeed, the second and third terms in Eq.(16a) are explicitly proportional to $p_\tau $, while the first term is also proportional to  $p_\tau $ via Eq.(17a). Thus all three terms are proportional to $p_\tau$  and vanish when $p_\tau =0$. 
In a general case, the emergent Hall voltage consists of two contributions. The first term on the right-hand side of Eq. (\ref{E_Hall}) is proportional to the spatial variation of the spin accumulation, and the second term on the right-hand side of Eq. (\ref{E_Hall}) is directly proportional to the emerging magnetic field.
\begin{figure}[t]
	\includegraphics[width=.95\columnwidth]{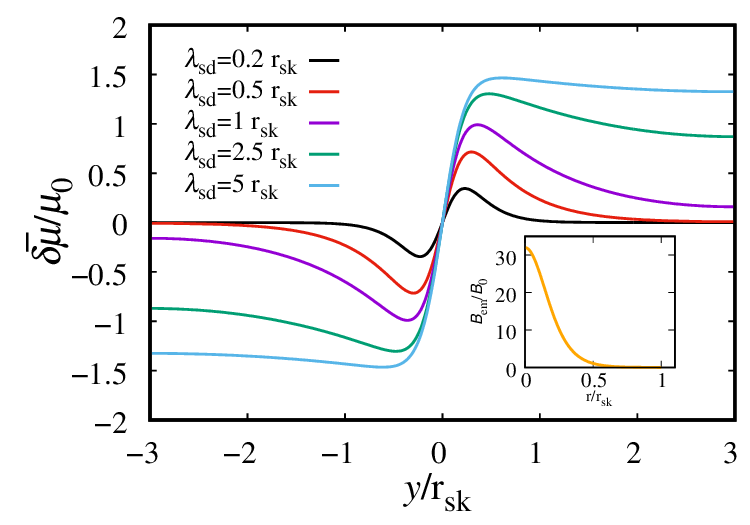}
	\caption{Spatial variation of the spin accumulation, generated by the skyrmion-induced real-space Berry curvature, Eq. (\ref{eq:delta_mu}), for different spin-diffusion lengths. The spin accumulation is normalized to $\mu_0=(e \tau E_x B_0 r_{\rm sk})/m_{\rm eff}$ and we set $w=3 r_{\rm sk}$. The inset shows the profile of the skyrmion-induced magnetic field, Eq. (\ref{eq:emergent_field}).} 
 \label{fig:mu}
\end{figure}

\begin{figure}[t]
	\includegraphics[width=.95\columnwidth]{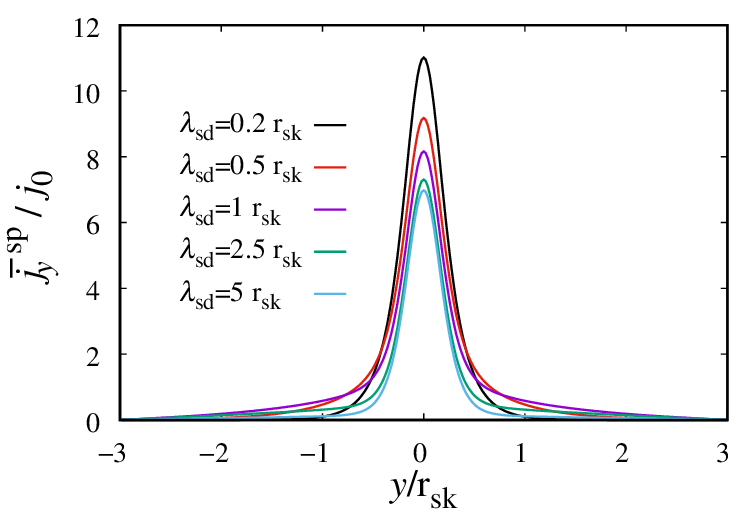}
	\caption{Spatial variation of the  topological spin Hall current density, Eq. (\ref{eq:j_sp2}), generated by the skyrmion-induced real-space Berry curvature, for several spin diffusion lengths. The spin current is normalized to $j_0=(1-p_{\tau}^2)(e \sigma_0 \tau {E_x} B_0 )/m_{\rm eff}$ and we set $w=3 r_{\rm sk}$. }  \label{fig:js}
\end{figure}
To find the spin current density and the Hall voltage, we need to know the spatial dependence of the average spin accumulation $\overline{\delta\mu}(y)$. Now, we should implement other boundary conditions. The spin current should be zero at both edges, $\overline{j}^{\rm sp}_{y}(\pm w)=0$. Using this boundary condition, together with Eqs. (\ref{eq:diffusion_sub_lattice1}) and (\ref{eq:j_sp}), we find the nonequilibrium spin accumulation profile as,  
{\small{
\begin{equation}
\begin{aligned}
\overline{\delta\mu}(y)=&\dfrac{e \tau {E_x}}{2m_{\mathrm{eff}}}\Bigg(\dfrac{\sinh(\frac{y}{\lambda_{\rm{sd}}})\exp(\dfrac{-w}{\lambda_{\rm{sd}}})}{\cosh(\dfrac{w}{\lambda_{\rm{sd}}})}\bigintssss_{-w}^{+w}\overline{B}_{\rm em}(\tilde{y})\exp(\dfrac{\tilde{y}}{\lambda_{\rm{sd}}})d\tilde{y}\\
&+\bigintssss_{-w}^{+w}\overline{B}_{\rm em}(\tilde{y})\dfrac{y-\tilde{y}}{|y-\tilde{y}|}\exp(\dfrac{-|y-\tilde{y}|}{\lambda_{\rm{sd}}})d\tilde{y}\Bigg),
\end{aligned}
\label{eq:delta_mu}
\end{equation}
}}
The first term on the right-hand side of this expression is the homogeneous solution of Eq. (\ref{eq:diffusion_sub_lattice1}), and the second term is its particular solution.

Having the spin accumulation, Eq. (\ref{eq:delta_mu}), we find the spin Hall current density from Eq. ~(\ref{eq:j_sp}), 

\begin{equation}
\begin{aligned}
& \overline{j}^{\rm sp}_{y}(y)=\sigma_0(1-p_{\tau}^2)\dfrac{e\tau E_x}{m_{\mathrm{\rm eff}}}\Bigg[ +\overline{B}_{\rm em}(y) \nonumber \\
& + \dfrac{1}{2\lambda_{\rm sd}}\Bigg( 
\bigintssss_{-w}^{+w}\overline{B}_{\rm em}(\tilde{y})\exp(-\dfrac{|y-\tilde{y}|}{\lambda_{\rm sd}})d\tilde{y}\\
&-\dfrac{\cosh(\frac{y}{\lambda_{\rm sd}})\exp(\dfrac{-w}{\lambda_{\rm sd}})}{\cosh(\dfrac{w}{\lambda_{\rm sd}})}\bigintssss_{-w}^{+w}\overline{B}_{\rm em}(\tilde{y})\exp(\dfrac{\tilde{y}}{\lambda_{\rm sd}})d\tilde{y}
\Bigg)\Bigg].
\end{aligned}
\label{eq:j_sp2}
\end{equation}

This expression describes the spatial dependence of the spin Hall current density in the AFM stripe in the presence of an emerging field of a skyrmion. This transverse spin current has two contributions. There is an extrinsic contribution arising from the spin accumulation gradient (the first two integrals) and an intrinsic contribution from the emergent magnetic field (the last term).
These two contributions, however, are not separable. Spin current  is created mainly within the skyrmion core (the term proportional to the emerging field), but it  is modified by the spin diffusion.  
Moreover, the interplay of the two terms terms in spin current leads to an irregular behavior outside the skyrmion core.

\section{Spin accumulation, spin current density, and Hall voltage} 
The spin accumulation and spin current density are calculated from Eqs. (\ref{eq:delta_mu}) and (\ref{eq:j_sp2}), respectively. The integrals cannot be calculated analytically and thus we integrate them numerically.
Figure \ref{fig:mu} shows the spatial variation of the nonequilibrium spin accumulation, Eq. (\ref{eq:delta_mu}), for various spin diffusion lengths. The inset shows the profile of the emerging magnetic fields, Eq. (\ref{eq:emergent_field}). 

The spin accumulation displays a nonmonotonic spatial behaviour, rising within the first half of the skyrmion radius but declining as it extends towards the edges. 
Beyond the skyrmion region, the spin accumulation gradually vanishes for shorter diffusion lengths, whereas it reaches a constant value for longer diffusion lengths.  
Reducing the spin diffusion length results in a decrease in the spin accumulation amplitude as one may expect.

The associated topological spin Hall current flowing across the stripe is presented in Fig. \ref{fig:js} for various spin diffusion lengths. The spin Hall current reduces dramatically outside the skyrmion core, and eventually vanishes outside the skyrmion. Inside the skyrmion core, the amplitude of the spin current increases with reducing the spin diffusion length, while it decreases away from the skyrmion core.

\begin{figure}[t]
\includegraphics[width=.95\columnwidth]{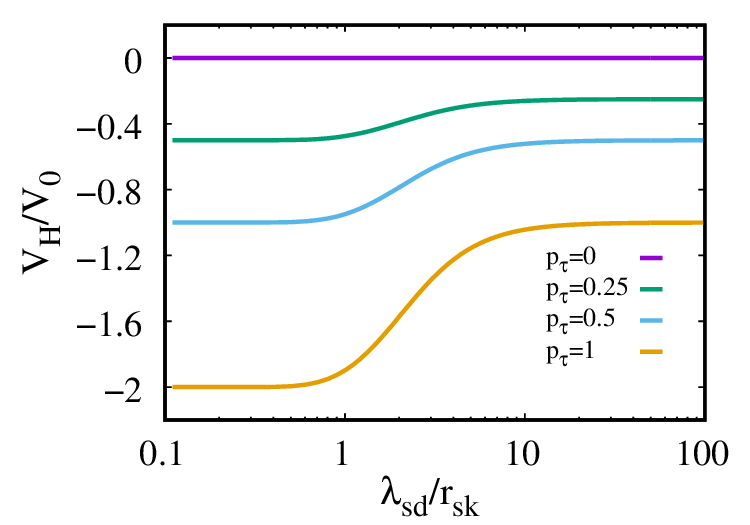}
\caption{The Hall voltage as a function of the spin diffusion length for indicated values of the spin asymmetry of the relaxation time. The Hall voltage is normalized to $V_0={h\tau E_x}/({2 L m_{\rm eff}})$.}  \label{fig:vh}
\end{figure}

Finally, in Fig. \ref{fig:vh} we plot the Hall voltage as a function of the spin diffusion length for the indicated values of the spin asymmetry parameter $p_{\tau}$. When the spin asymmetry of the relaxation time increases, the amplitude of the Hall voltage increases as well. In turn, an increase in the spin diffusion length leads to a decrease in the Hall voltage.

\section{Summary and concluding remarks} 
We computed the topological charge and spin Hall effects, as well as the spin accumulation, in a compensated AFM system, arising from a real-space Berry curvature induced by a skyrmion. We calculated these quantities in the diffusive regime using the  semiclassical Boltzmann formalism. Our model describes a single skyrmion in either a metallic AFM thin film or at the interface of an AFM insulator and a metal.
We considered both spin-dependent scattering and spin-flip mechanisms in our calculations.
We found both intrinsic and extrinsic contributions to the spin Hall effect and spin accumulation, and we showed that the spin Hall current vanishes rapidly outside the skyrmion. On the other hand, the spin accumulation, which is a measurable quantity, can be finite in systems with the spin diffusion length larger than the skyrmion size.  
In addition, we showed that the Hall voltage vanishes in the absence of spin asymmetry in relaxation times,  $p_\tau =0$, and can be finite when such an asymmetry appears, $p_\tau \ne 0$, for instance due to scattering on magnetic defects. 
Direct detection of skyrmions is a challenge because of the absence of net magnetization in compensated AFM systems. We argue that the electrical detection of AFM skyrmions is possible by measuring the spin accumulation and/or Hall voltage. 

\section*{Acknowledgements}
This work has been supported by the Norwegian Financial Mechanism 2014- 2021 under the Polish – Norwegian Research Project NCN GRIEG “2Dtronics” no. 2019/34/H/ST3/00515.


\begin{thebibliography}{10}
\expandafter\ifx\csname url\endcsname\relax
  \def\url#1{\texttt{#1}}\fi
\expandafter\ifx\csname urlprefix\endcsname\relax\def\urlprefix{URL }\fi
\expandafter\ifx\csname href\endcsname\relax
  \def\href#1#2{#2} \def\path#1{#1}\fi

\bibitem{PhysRevLett.78.3773}
G. Tatara and H. Fukuyama, 
\href {https://link.aps.org/doi/10.1103/PhysRevLett.78.3773}{Phys. Rev. Lett. 78  (1997) 3773.}

\bibitem{PhysRevLett.79.5110}
P. M. Levy and S. Zhang, 
\href {https://link.aps.org/doi/10.1103/PhysRevLett.79.5110}{Phys. Rev. Lett. 79 (1997) 5110.}


\bibitem{PhysRevB.59.138}
J. B. A. N. van Hoof, K. M. Schep, A. Brataas, G. E. W. Bauer, and P. J. Kelly, 
\href {https://link.aps.org/doi/10.1103/PhysRevB.59.138}{Phys. Rev. B 59 (1999) 138.} 

\bibitem{PhysRevB.60.3406}
A. Brataas, G. Tatara, and G. E. W.  Bauer, 
\href {https://link.aps.org/doi/10.1103/PhysRevB.60.3406}{Phys. Rev. B 60 (1999) 3406}.

\bibitem{PhysRevB.63.224412}
E. \ifmmode \check{S}\else \v{S}\fi{}im\'anek, 
\href {https://link.aps.org/doi/10.1103/PhysRevB.63.224412}{Phys. Rev. B 63 (2001) 224412}.

\bibitem{10.1063/1.1578165}
M. Tsoi, R. E. Fontana, and S. S. P. Parkin,
\href {https://doi.org/10.1063/1.1578165}{Appl. Phys. Lett. 83 (2003) 2617-2619}.

\bibitem{Yamanouchi2004}
M. Yamanouchi, D. Chiba, F. Matsukura, and H. Ohno,
\href {https://doi.org/10.1038/nature02441}{Nature 428 (2004) 539–542}.

\bibitem{PhysRevLett.92.086601}
G. Tatara and H. Kohno,
\href {https://link.aps.org/doi/10.1103/PhysRevLett.92.086601}{Phys. Rev. Lett. 92 (2004)  086601}.


\bibitem{TATARA2008213}
G. Tatara, H. Kohno, and J. Shibata,
\href {https://www.sciencedirect.com/science/article/pii/S0370157308002597}{Phys. Rep. 468 (2008) 213-301}.



\bibitem{Swaving_2012}
A. C. Swaving and R. A. Duine,
\href {https://dx.doi.org/10.1088/0953-8984/24/2/024223}{J. Phys. Condens. Matter 24 (2011) 024223}.

\bibitem{PhysRevLett.110.127208}
E. G. Tveten, A. Qaiumzadeh, O. A. Tretiakov, and A. Brataas,
\href {https://link.aps.org/doi/10.1103/PhysRevLett.110.127208}{Phys. Rev. Lett. 110 (2013) 127208}.

\bibitem{PhysRevB.93.180408}
Y. Yamane, J. Ieda, and J. Sinova, 
\href {https://link.aps.org/doi/10.1103/PhysRevB.93.180408}{Phys. Rev. B 93 (2016) 180408}.

\bibitem{PhysRevB.96.165303}
Y. Araki and K. Nomura, 
\href {https://link.aps.org/doi/10.1103/PhysRevB.96.165303}{Phys. Rev. B 96 (2017) 165303}.

\bibitem{PhysRevB.102.184413}
J. H. Zheng, A. Brataas, M. Kl\"aui, and A. Qaiumzadeh,
\href {https://link.aps.org/doi/10.1103/PhysRevB.102.184413}{Phys. Rev. B 102 (2020) 184413}.

\bibitem{Liang2015}
D. Liang, J. P. DeGrave, M. J. Stolt, Y. Tokura, and S. Jin,
\href {https://doi.org/10.1038/ncomms9217}{Nat. Commun. 6 (2015) 8217 }.

\bibitem{WANG2022100971}
H. Wang, Y. Dai,  G. M. Chow, and J. Chen,
\href {https://www.sciencedirect.com/science/article/pii/S0079642522000524}{Prog. Mater. Sci. 130 (2022) 100971}.


\bibitem{nphys2231}
T. Schulz et al.
\href {https://doi.org/10.1038/nphys2231}{Nat. Phys. 8 (2011) 136804}.

\bibitem{nnano.2013.243}
N. Nagaosa and Y.Tokura, 
\href {https://doi.org/10.1038/nnano.2013.243}{Nat. Nanotech. 8 (2013) 899–911}.

\bibitem{ncomms2442}
J. Iwasaki, M. Mochizuki, and N. Nagaosa, 
\href {https://doi.org/10.1038/ncomms2442}{Nat. Commun. 4 (2013) 1463}.


\bibitem{Litzius2017}
K. Litzius et al. 
\href {https://doi.org/10.1038/nphys4000}{Nat. Phys. 13 (2017) 170-175}.

\bibitem{nphys3883}
W. Jiang et al. 
\href {https://doi.org/10.1038/nphys3883}{Nat. Phys. 13 (2017) 162-169}.

\bibitem{PhysRevB.95.094401}
I. A. Ado,  O. A. Tretiakov, and M. Titov, 
\href {https://link.aps.org/doi/10.1103/PhysRevB.95.094401}{Phys. Rev. B. 95 (2017) 094401}.

\bibitem{PhysRevLett.100.127204}
O. A. Tretiakov, D. Clarke, G.-Wei Chern, Y. B. Bazaliy, and O. Tchernyshyov, 
\href {https://link.aps.org/doi/10.1103/PhysRevLett.100.127204}{Phys. Rev. Lett. 100 (2008) 127204}.

\bibitem{PhysRevB.78.134412}
D. J. Clarke, O. A. Tretiakov, G.-W. Chern, Ya. B. Bazaliy, and O. Tchernyshyov, 
\href {https://link.aps.org/doi/10.1103/PhysRevB.78.134412}{Phys. Rev. B 78, 134412 (2008)}. 

\bibitem{PhysRevLett.107.136804}
J. Zang, M. Mostovoy, J. Han, and N. Nagaosa, 
\href {https://link.aps.org/doi/10.1103/PhysRevLett.107.136804}{Phys. Rev. Lett.  107 (2011) 136804}.

\bibitem{Hans2012}
H. B. Braun,
\href {https://doi.org/10.1080/00018732.2012.663070}{Adv. Phys.  61 (2012) 1-116}.

\bibitem{Tomasello2014}
R. Tomasello et al.
\href {https://doi.org/10.1038/srep06784}{Sci. Rep. 4 (2014) 6784}.



\bibitem{PhysRevLett.30.230}
A. A. Thiele,
\href {https://link.aps.org/doi/10.1103/PhysRevLett.30.230}{Phys. Rev. Lett. 30 (1973) 230}.

\bibitem{Martinez_2016}
J. C. Martinez and M. B. A. Jalil,
\href {https://dx.doi.org/10.1088/1367-2630/18/3/033008}{New J. Phys. 18 (2016) 033008}.

\bibitem{PhysRevLett.83.3737}
J. Ye, Y. B. Kim, A. J. Millis, B. I. Shraiman, P. Majumdar, and Z. Te\ifmmode \check{s}\else \v{s}\fi{}anovi\ifmmode \acute{c}\else \'{c}\fi{},
\href {https://link.aps.org/doi/10.1103/PhysRevLett.83.3737}{Phys. Rev. Lett. 83 (1999) 3737}.

\bibitem{PhysRevLett.93.096806}
P. Bruno, V. K. Dugaev, and M. Taillefumier,
\href {https://link.aps.org/doi/10.1103/PhysRevLett.93.096806}{Phys. Rev. Lett. 93 (2004) 096806}.

\bibitem{PhysRevLett.102.186602}
A. Neubauer, C. Pfleiderer, B. Binz, A. Rosch, R. Ritz, P. G. Niklowitz, and P. B\"oni,
\href {https://link.aps.org/doi/10.1103/PhysRevLett.102.186602}{Phys. Rev. Lett. 102 (2009) 186602}.


\bibitem{PhysRevLett.102.186601}
L. Minhyea, W. Kang, Y. Onose, Y. Tokura, and N. P. Ong,
\href {https://link.aps.org/doi/10.1103/PhysRevLett.102.186601}{Phys. Rev. Lett. 102 (2009) 186601}.

\bibitem{PhysRevB.91.245115}
Y. Ohuchi, Y. Kozuka, M. Uchida, K. Ueno, A. Tsukazaki, and M. Kawasaki,
\href {https://link.aps.org/doi/10.1103/PhysRevB.91.245115}{Phys. Rev. B 91 (2015) 245115}.

\bibitem{PhysRevLett.117.027202}
K. S. Denisov, I. V. Rozhansky, N. S. Averkiev, and E. L\"ahderanta,
\href {https://link.aps.org/doi/10.1103/PhysRevLett.117.027202}{Phys. Rev. Lett. 117 (2016) 027202}.


\bibitem{PhysRevB.95.064426}
P. B. Ndiaye, C. A. Akosa, and A. Manchon,
\href {https://link.aps.org/doi/10.1103/PhysRevB.95.064426}{Phys. Rev. B 95 (2017) 064426}.

\bibitem{PhysRevB.97.134401}
S. S.-L. Zhang and O.-L. Heinonen, 
\href {https://link.aps.org/doi/10.1103/PhysRevB.97.134401}{Phys. Rev. B. 97 (2018) 134401}.

\bibitem{PhysRev.95.1154}
K. Robert and J. M. Luttinger,
\href {https://link.aps.org/doi/10.1103/PhysRev.95.1154}{ Phys. Rev. 95 (1954) 1154 }.

\bibitem{RevModPhys.82.1539}
N. Nagaosa, J. Sinova, S. Onoda, A. H. MacDonald, and N. P. Ong,
\href {https://link.aps.org/doi/10.1103/RevModPhys.82.1539}{ Rev. Mod. Phys. 82 (2010) 1539}.


\bibitem{PhysRevLett.112.017205}
H. Chen, Q. Niu, and A. H. MacDonald,
\href {https://link.aps.org/doi/10.1103/PhysRevLett.112.017205}{ Phys. Rev. Lett. 112 (2014) 017205}.

\bibitem{PhysRevLett.98.246601}
S. E. Barnes and S. Maekawa, 
\href {https://link.aps.org/doi/10.1103/PhysRevLett.98.246601}{Phys. Rev. Lett. 98 (2007) 246601}.

\bibitem{PhysRevB.64.104411}
V. K. Dugaev, A. Cr\'epieux, and P. Bruno,
\href {https://link.aps.org/doi/10.1103/PhysRevB.64.104411}{ Phys. Rev. B 64 (2001) 104411}.

\bibitem{PhysRevB.64.014416}
A. Cr\'epieux and P. Bruno,
\href {https://link.aps.org/doi/10.1103/PhysRevB.64.014416}{Phys. Rev. B 64 (2001) 014416}.

\bibitem{PhysRevB.68.045327}
D. Culcer, A. MacDonald, and Q. Niu,
\href {https://link.aps.org/doi/10.1103/PhysRevB.68.045327}{ Phys. Rev. B 68 (2003) 045327}.

\bibitem{APL1.4967006}
C. Jin, C. Song, J. Wang, and Q. Liu, 
\href {https://doi.org/10.1063/1.4967006}{Appl. Phys. Lett. 109 (2016) 182404}.

\bibitem{PhysRevLett.116.147203}
J. Barker and O. A. Tretiakov, 
\href {https://link.aps.org/doi/10.1103/PhysRevLett.116.147203}{Phys. Rev. Lett. 116 (2016) 147203}.

\bibitem{Zhang2016_1}
X. Zhang, Y. Zhou, and M. Ezawa,\href {https://doi.org/10.1038/ncomms10293}{ Nat. Commun. 7 (2016) 10293}.

\bibitem{Zhang2016}
X. Zhang, Y. Zhou, and M. Ezawa,\href {https://doi.org/10.1038/srep24795}{ Sci. Rep. 6 (2016) 24795}.

\bibitem{PhysRevB.99.054423}
R. Khoshlahni, A. Qaiumzadeh, A. Bergman, and A. Brataas,
\href {https://link.aps.org/doi/10.1103/PhysRevB.99.054423}{Phys. Rev. B 99  (2019) 054423}.

\bibitem{TRETIAKOV2021333}
O. A. Tretiakov,
\href {https://www.sciencedirect.com/science/article/pii/B9780128208151000092}{Woodhead Publishing Series in Electronic and Optical Materials (2021) 333-345}.

\bibitem{GOBEL20211}
B. G\"obel, I. Mertig, and O. A. Tretiakov,
\href {https://www.sciencedirect.com/science/article/pii/S0370157320303525}{Phys. Rep. 895 (2021) 1-28}.

\bibitem{Amin2023}
O. J. Amin et al.
\href {https://doi.org/10.1038/s41565-023-01386-3}
{Nat. Nanotechnol. 18 (2023) 849–853 }.
\bibitem{PhysRevB.86.245118}
R. Cheng and Q. Niu, 
\href {https://link.aps.org/doi/10.1103/PhysRevB.86.245118}{Phys. Rev. B 86 (2012) 245118}.


\bibitem{RRL1700007}
 P. M. Buhl, F. Freimuth, S. Bl{\"u}gel, and Y. Mokrousov,
\href {https://doi.org/10.1002/pssr.201700007}{ Phys. Status Solidi RRL 11 (2017) 1700007}.

\bibitem{PhysRevB.96.060406}
B. G\"obel, A. Mook, J. Henk, and I. Mertig, 
\href {https://link.aps.org/doi/10.1103/PhysRevB.96.060406}{Phys. Rev. B 96 (2017) 060406}.

\bibitem{PhysRevLett.121.097204}
C. A. Akosa, O. A. Tretiakov, G. Tatara, and A. Manchon, 
\href {https://link.aps.org/doi/10.1103/PhysRevLett.121.097204}{Phys. Rev. Lett. 121 (2018) 097204}.

\bibitem{nakazawa2023topological}
K. Nakazawa, K. Hoshi, J. J. Nakane, J.-I. Ohe, and H. Kohno, 
\href {https://doi.org/10.48550/arXiv.2304.02850}{ 	arXiv:2304.02850 (2023)}.

\bibitem{Gao}
S. Gao et al,  
\href {https://doi.org/10.1038/s41586-020-2716-8}{ Nature 586 (2020) 37 }.

\bibitem{Jani}
H. Jani et al,  
\href {https://doi.org/10.1038/s41586-021-03219-6}{ Nature 590 (2021) 74 }.

\bibitem{Legrand}
W. Legrand et al,  
\href {https://doi.org/10.1038/s41563-019-0468-3}{ Nature Mater. 19 (2020) , 34}.

\bibitem{Qiang_2022}
 B. Qiang et al. 
\href {https://dx.doi.org/10.35848/1347-4065/ac9a92} { Jpn. J. Appl. Phys. 61 (2022) 120901}.

\bibitem{PhysRevB.95.134424}
H. B. M. Saidaoui, X. Waintal, and A. Manchon,
\href {https://link.aps.org/doi/10.1103/PhysRevB.95.134424}{Phys. Rev. B 95 (2017) 134424}.

\bibitem{PhysRevLett.120.197202}
A. Qaiumzadeh, I. A. Ado,  R. A. Duine,  M. Titov, and A. Brataas, 
\href {https://link.aps.org/doi/10.1103/PhysRevLett.120.197202}{Phys. Rev. Lett. 120 (2018) 197202}.

\bibitem{PhysRevResearch.5.L022065}
M. M. S. Barbeau, M. Titov, M. I. Katsnelson, and A. Qaiumzadeh, 
\href {https://link.aps.org/doi/10.1103/PhysRevResearch.5.L022065}{Phys. Rev. Research 5 (2023) L022065}.

\bibitem{PhysRevB.98.195439}
K. S. Denisov, I. V. Rozhansky, N. S. Averkiev, and E. L\"ahderanta, 
\href {https://link.aps.org/doi/10.1103/PhysRevB.98.195439}{Phys. Rev. B 98 (20) 195439}.

\bibitem{Volovik1987}
G. E. Volovik, 
\href {https://dx.doi.org/10.1088/0022-3719/20/7/003}{J. Phys. C 20 (1987)  L83}.

\bibitem{PhysRevB.77.134407}
Y. Tserkovnyak and M. Mecklenburg, 
\href {https://link.aps.org/doi/10.1103/PhysRevB.77.134407}{Phys. Rev. B. 77 (2008) 134407}.


\bibitem{PhysRevLett.102.086601}
S. Zhang and S. S.-L. Zhang, 
\href {https://link.aps.org/doi/10.1103/PhysRevLett.102.086601}{Phys. Rev. Lett. 102 (2009) 086601}.


\bibitem{akosa1}
C. A. Akosa, {\it{Spin Transport in Ferromagnetic and Antiferromagnetic Textures}},  PhD thesis,
\href {https://repository.kaust.edu.sa/handle/10754/621988?show=full}{King Abdullah University of Science and Technology (KAUST), Thuwal, Saudi Arabia (2016)}.


\bibitem{Ashcroft}
N. W. Ashcroft and N. D. Mermin, {\it{Solid State Physics}}
\href {}{Harcourt College Publishers, San Diego (1975)}.

\bibitem{PhysRevB.67.052407}
Y. Qi and S. Zhang, 
\href {https://link.aps.org/doi/10.1103/PhysRevB.67.052407}{Phys. Rev. B 67 (2003) 052407}.

\bibitem{PhysRevB.82.184423}
S. S.-L. Zhang and S. Zhang,  
\href {https://link.aps.org/doi/10.1103/PhysRevB.82.184423}{Phys. Rev. B 82 (2010) 184423}.

\bibitem{valet}
T. Valet and A. Fert, 
\href {https://link.aps.org/doi/10.1103/PhysRevB.48.7099}{Phys. Rev. B 48 (1993) 7099}.

\bibitem{10.1063/1.356868}
A. Fert, T. Valet, and J. Barna\'s, 
\href {https://doi.org/10.1063/1.356868}{Appl. Phys. 75  (1994) 6693}.


\end{thebibliography}
\end{document}